\documentclass[12pt]{article}
\usepackage{graphicx}
\usepackage{epstopdf}
\usepackage{subfigure}
\usepackage{color}
\usepackage{listings,booktabs}
\usepackage{amssymb,amsmath}
\newcommand{\be}{\begin{equation}}
\newcommand{\ee}{\end{equation}}
\newenvironment{Eqnarray}%
     {\arraycolsep 0.14em\begin{eqnarray}}{\end{eqnarray}}
\newcommand{\ba}{\begin{Eqnarray}}
\newcommand{\ea}{\end{Eqnarray}}

\title{Suppression of the Higgs Dimuon Decay}
\date{}
\begin{document}
\author{
\large
P.~M.~Ferreira,$^{(1,2)}$\thanks{E-mail: ferreira@cii.fc.ul.pt} \
and Marc Sher$^{(3)}$\thanks{E-mail: mtsher@wm.edu} \
\\*[3mm]
\small $^{(1)}$ Instituto Superior de Engenharia de Lisboa,
\small 1959-007 Lisboa, Portugal
\\*[2mm]
\small $^{(2)}$ Centro de F\'\i sica Te\'orica e Computacional,
University of Lisbon,
\small 1649-003 Lisboa, Portugal
\\*[2mm]
\small $^{(3)}$ High Energy Theory Group, William \& Mary,
\small Williamsburg, Virginia 23187, U.S.A.
}
\date{\today}

\maketitle

\begin{abstract}
It is often stated that elimination of tree-level flavor-changing neutral currents in multi-Higgs models
requires that all fermions of a given charge to couple to the same Higgs boson.   A counterexample was
provided by Abe, Sato and Yagyu (ASY) in a muon-specific two-Higgs doublet model.
In this model, all fermions except the muon couple to one Higgs and the muon couples to the
other.
We study the phenomenology of the model and show that there is a wide range of parameter-space
in which the branching ratios of the 125 GeV Higgs are very close to their Standard Model values, with
the exception of the branching ratio into muons, which can be substantially suppressed -- this is an
interesting possibility, since the current value of this branching ratio is $0.5 \pm 0.7$ times the
Standard Model value.
We also study the charged Higgs boson and show that, if it is lighter than $200$ GeV, it could
have a large branching ratio into $\mu\nu$ - even substantially larger than the usual decay into
$\tau\nu$.   The decays of the heavy neutral scalars are also studied.   The model does have a
relationship between the branching ratios of the 125 GeV Higgs into $Z$'s, $\tau$'s and $\mu$'s, which can be tested
in future accelerators.
\end{abstract}

\section{Introduction}

The Higgs boson was initially discovered \cite{Aad:2012tfa,Chatrchyan:2012xdj} through its decay
into gauge bosons.    Since then, the coupling of the Higgs to third generation fermions has also
been determined with increasing accuracy \cite{Aad:2015vsa, Chatrchyan:2014nva, Aaboud:2018zhk,
Sirunyan:2018kst, Sirunyan:2018hoz, Aaboud:2018urx}.     However, the coupling to second generation
fermions has not yet been observed.    CMS and ATLAS have performed \cite{Sirunyan:2018hbu,ATLAS:2019ain}
a search for the dimuon decay; the most recent study by ATLAS \cite{ATLAS:2019ain} finds a branching
ratio of $0.5\pm 0.7$ times the Standard Model branching ratio (the uncertainty is one standard
deviation).     While this is certainly consistent with the Standard Model, it leads one to wonder
what the consequences would be if the dimuon decay is not discovered in the near future, implying
that the branching ratio is substantially below that of Standard Model.

Since the Higgs branching ratio for $\tau$ pairs has been observed to be fairly close to the
Standard Model value, a
nondiscovery of the dimuon decay would imply that the interaction of the Higgs boson with charged
leptons does not follow SM expectations -- the Higgs would couple to different flavours in a way not
simply proportional to their masses, regardless of generations. Thus (apart from their mass differences) the
second and third generations would not be just replicas of each other, the Higgs interactions with each
would follow different rules.
A general discussion of models in which new physics at a high scale generates the light generation masses
 appeared in Botella, et al. \cite{Botella:2016krk}.   As they point out, the simplest model would be the
 addition of a Higgs boson which does not couple to the third generation.   A well-studied set of models that
 have this property are the Branco, Grimus and Lavoura models \cite{Branco:1996bq,Botella:2009pq}.   An
 interesting feature of these models is that they contain tree-level flavor-changing neutral currents (FCNC),
 but these are related directly to elements of the CKM (or PMNS) matrix.

FCNC have not been detected, and so one can consider a model in which there are no tree-level FCNC, but
the muon and tau leptons couple to different Higgs bosons.   Such a model, called the muon-specific Two
Higgs Doublet (2HDM) model, was developed by Abe, Sato and Yagyu \cite{Abe:2017jqo} (ASY).   They use a
$Z_4$ symmetry, under which the muon and tau have different quantum numbers, and break this softly.  Ivanov and Nishi have pointed out \cite{Ivanov:2013bka} that the actual symmetry group of the model is a softly broken $Z_2$ with a $U(1)$ corresponding
to muon number (this does not affect ASY's results, but is more precise).  The
model has no tree-level FCNC and the Yukawa couplings for the muon and tau are no longer simply
proportional to their masses with the proportionality coefficient being the same for all flavours:
rather, the ASY model can substantially enhance or suppress the muon interactions of scalars relative
to those with tau leptons. The purpose
of their model was to attempt an explanation of the muon g-2 anomaly, and for the parameters they considered
the dimuon coupling of the 125 GeV Higgs is not suppressed.   Their model can address the g-2 anomaly, but as
we will see, it requires a very narrow region of parameter-space.

In this paper, we will consider the muon-specific model without requiring that it also address the muon g-2
 anomaly.   This gives a much wider region of parameter-space, and we will see that the coupling of the
 125 GeV Higgs to muons can be easily suppressed, without suppressing the coupling to tau pairs.
 The model is introduced in Section 2, and the parameters chosen by ASY will be discussed in Section 3.
  In Section 4, we study the phenomenology of the $125$ GeV Higgs, the charged Higgs and the heavy neutral
  Higgs bosons.  Section 5 contains our conclusions.

\section{The Muon Specific 2HDM}

The ASY muon specific 2HDM uses a $Z_4$ discrete symmetry.    Here, we present their model, following their
work closely.    As with other 2HDMs with a discrete symmetry, one Higgs doublet , $\Phi_2$, has quantum
number +1 and the other, $\Phi_1$, has quantum number -1.    All fermions except the second generation 
leptons have $Z_4$ quantum number +1.    Thus, $\Phi_1$ does not couple to these fermions; for them, 
the model is  similar to a type I 2HDM  (see Ref. \cite{Branco:2011iw} for a detailed review).     
The $Z_4$ quantum number of the right-handed muon, and of the second generation left leptonic doublet, 
is $i$, and thus there is a coupling of $\Phi_1$ to the muons (see Table I of ref~\cite{Abe:2017jqo} for 
the $Z_4$ charges of all fields in the model).

 As noted earlier, Ivanov and Nishi\cite{Ivanov:2013bka} showed that the actual symmetry group of the
 model is a softly broken $Z_2$ (as in the usual type I model), and a global $U(1)$ corresponding to
 muon number.    The fields odd under the $Z_2$ are the $\Phi_1$ and the $\mu_R$, while the $U(1)$
 quantum numbers vanish for all fields other than the left and right handed muons.    This result is
 precisely the same Lagrangian as the ASY model, but is clearly a larger symmetry (in fact, replacing
 the $Z_4$ with any $Z_N$ where N is even and greater than 2 yields the same model).    This doesn't
 affect the ASY model since the Lagrangian is the same.

 The Yukawa Lagrangian involving leptons is
\begin{equation}
{\cal L} = -\bar{L}_L \Phi_1 Y_1 E_R -\bar{L}_L \Phi_2 Y_2 E_R + {\rm h.c.}
\end{equation}
The $Y_1$ and $Y_3$ are $3\times 3$ matrices in flavor space.   Defining the left-handed (right-handed) lepton field $L_L$ ($E_R$) as
\begin{equation}
L_L = (\ell^e_L,\ell^\mu_L,\ell^\tau_L)^T,\qquad E_R = (e_R,\mu_R,\tau_R)^T,
\end{equation}
the $Z_4$ symmetry gives the lepton Yukawa matrices as
\begin{align}
Y_1 =\begin{pmatrix}0& 0& 0\\ 0& y_\mu& 0\\ 0& 0& 0\end{pmatrix},\quad Y_2 =\begin{pmatrix} y_e& 0& y_{e\tau}\\ 0& 0& 0\\ y_{\tau e}& 0& y_\tau\end{pmatrix}.
\end{align}
Since these matrices commute, they are simultaneously diagonalizable and thus there are no tree-level FCNC.
The off-diagonal terms in $Y_2$ can be set to zero by field rotations.

The Higgs potential is the same as in the usual 2HDM with a softly broken $Z_2$ symmetry. The potential may be written as
\begin{align}
 V &= m^2_{11}|\Phi_1|^2+m^2_{22}|\Phi_2|^2
 +m^2_{12}\left[\Phi_1^\dagger\Phi_2+{\rm h.c.}\right] \nonumber \\
  &\,\,+\frac{\lambda_1}{2}|\Phi_1|^4
   +\frac{\lambda_2}{2}|\Phi_2|^4 +\lambda_3 |\Phi_1|^2 |\Phi_2|^2   +\lambda_4 |\Phi_1^\dagger\Phi_2|^2
   +\frac{\lambda_5}{2}\left[\left(\Phi_1^\dagger\Phi_2\right)^2+{\rm h.c.}\right]\,,
\label{eq:pot}
\end{align}
with all $8$ parameters real. Notice that though the lagrangian possesses a $Z_4$ symmetry (in fact
a $Z_2\times U(1)$ one), the ASY
transformation law for the scalar fields is simply $\Phi_1 \rightarrow -\Phi_1$ and
$\Phi_2 \rightarrow \Phi_2$, therefore it is not surprising that the form of the potential is identical
to that of the usual $Z_2$ symmetry considered in the 2HDM. We denote the VEVs of the Higgs fields by $v_1$
and $v_2$, and follow the usual convention in defining $\tan\beta\equiv v_2/v_1$.
The gauge eigenstates of the two neutral scalars are rotated into the mass eigenstates, as in the usual
convention, by a rotation angle $\alpha$.   Thus, the rotation from the Higgs basis (in which only one field
gets a VEV) to the mass basis is through an angle $\beta-\alpha$.  We will refer to $\sin(\beta-\alpha)$
($\cos(\beta-\alpha)$) as $s_{\beta\alpha}$ ($c_{\beta\alpha}$) respectively, and will occasionally refer to
$\tan\beta$ as $t_\beta$.    The expressions for the parameters of the scalar potential in terms of $\alpha$,
$\beta$, $v$, the masses of the physical scalar fields $h, H, A, H^\pm$ and the soft breaking parameter
are given in the ASY paper
\cite{Abe:2017jqo}\footnote{There is a typo in their equation 2.18.  The penultimate term should also be
multiplied by the expression in parenthesis in the last term.  This doesn't affect their work at all.}.

With these definitions, ASY write the interaction terms involving the $\tau$ and $\mu$ as (with $h$ ($H$)
being the lighter (heavier) scalar Higgs)
\begin{align}
{\cal L}_{\rm int} &=
-\frac{m_\tau}{v}\left[ (s_{\beta\alpha} + \frac{c_{\beta\alpha}}{t_\beta})\bar{\tau}\tau h + (c_{\beta\alpha} + \frac{s_{\beta\alpha}}{t_\beta})\bar{\tau}\tau H - i\frac{1}{t_\beta}\bar{\tau}\gamma_5\tau A\right]
\label{eq:yf}\\
&- \frac{m_\mu}{v}\left[ (s_{\beta\alpha}-t_\beta c_{\beta\alpha})\bar{\mu}\mu h + (c_{\beta\alpha}-t_\beta s_{\beta\alpha})\bar{\mu}\mu H + it_{\beta}\bar{\mu}\gamma_5\mu A\right]
\label{eq:ymu}\\
&-\frac{\sqrt{2}}{v}\left[\frac{m_\tau}{t_\beta}\bar{\nu}_\tau P_R\tau H^+ -  m_\mu t_\beta \bar{\nu}_\mu P_R\tau H^ + + {\rm h.c.}\right].
\label{eq:ych}
\end{align}
Here, $P_R$ is the right-handed projection operator.  Note that the muon couplings to the additional Higgs bosons are enhanced by a factor of $\tan\beta$.    Note also that the coupling of the $\tau$ to the $125$ GeV Higgs, $h$, is the same as the usual type I 2HDM coupling (in which the ratio of the coupling to that of the SM is $\cos\alpha/\sin\beta$).

\section{The Large $\tan\beta$ Limit and g-2 of the Muon}

As noted earlier, the muon specific 2HDM was first proposed by Abe, Sato and Yagyu (ASY) in Ref.
\cite{Abe:2017jqo}.    The purpose of their work was to use the model to explain the muon g-2 anomaly.
This is achieved by considering charged Higgs loops, whose coupling to muons is enhanced by $\tan\beta$.
Extremely large values of $\tan\beta$ are needed, typically of O(1000).     Normally, this large a value
would cause concern with perturbation theory, unitarity, electroweak precision observables, etc.
However, ASY show that these concerns will be alleviated if one chooses the free parameters of the model
carefully.    In particular, they require $s_{\beta\alpha} =1$ and a specific value for the soft symmetry
 breaking term $m_{12}^2$.   Note that the choice of $s_{\beta\alpha}=1$ makes the coupling of $h$ to the
 leptons identical to their Standard Model values for all $\tan\beta$.

The coupling of the $125$ GeV Higgs to muon pairs is the Standard Model coupling times
$s_{\beta\alpha} - t_\beta c_{\beta\alpha}$.   From the ATLAS result \cite{ATLAS:2019ain} which says
 that the 95\% upper limit on the branching ratio of $h\to\mu\mu$ is 1.7 times the Standard Model
 branching ratio, one concludes that $|s_{\beta\alpha} - t_\beta c_{\beta\alpha}|$ must be less than 1.3.
    For $\tan\beta=1000$, this means that $c_{\beta\alpha}$ is between $-0.0003$ and $0.0021$, or
 $s_{\beta\alpha} > 0.999998$.   This is, of course, consistent with their assumptions, but does seem
 highly fine-tuned.   In addition, the soft symmetry breaking term is also highly fine-tuned in their work.
 Note that between the two bounds for $c_{\beta\alpha}$, the coupling of the $125$ GeV Higgs to muon
 pairs could be smaller than the Standard Model value, which is the objective of this study.
 Since it is fine-tuned, we will ignore the issue of the muon g-2 anomaly and will consider values
 of $\tan\beta$ between 1 and 10, which will not require much fine-tuning.

\section{Numerical Analysis}

With the Lagrangian above, and the Higgs-quark-quark couplings the same as those for the Type I 2HDM,
one can calculate production cross sections and branching ratios. A preliminary scan showed that the model
can accomodate very large masses for the extra scalars (above 1 TeV) and remain compatible with LHC results --
this was to be expected, since the presence of the soft-breaking term $m^2_{12}$ in the potential of
Eq.~\ref{eq:pot} allows the model to have a decoupling regime. We have verified, however, that the more
interesting phenomenology of this muon-specific model occurs for lower masses of the extra scalars, which is
the reason that informs the scans we will now present: to scan the parameter-space, we
randomly generate extra scalar masses in the range (all mass units in GeV):
$130 < m_H < 500, 100 < m_A , m_{H^+} < 500$ and consider the ranges
 $1 < \tan\beta <10, 0.01 < |\lambda_5| < 4\pi, {\rm and}\ 0.9 \leq \sin(\beta-\alpha) \leq 1$.
 $\lambda_5$ is the quartic parameter in the 2HDM potential of Eq.~\ref{eq:pot}, and its range of variation
 was chosen to maximize the efficiency of the scan, after initial trial runs (this range favours
 compliance with unitarity conditions, for instance).
 The lower limit on the charged Higgs mass satisfies the bounds coming from direct searches of this
 particle. We have also chosen $h$ to be the SM-like, with mass 125 GeV, scalar observed at LHC, and $H$ the
 heavier CP-even scalar of the model~\footnote{It is still possible, though strongly constrained, to have $h$
 be the heavier CP-even scalar~\cite{Ferreira:2012my}. This possibility was used, for instance, to try to
 account for a possible excess in the diphoton channel at 96 GeV~\cite{Biekotter:2019kde}, though
 in that work the the 2HDM needed to be complemented with a real singlet (the N2HDM).}.
 We check that unitarity, boundedness from below and electroweak constraints on the quantities $S$ and $T$
are satisfied. The generated
 parameters, as well as the chosen quark couplings
ensure that constraints from $b$ physics (such as the $Z\rightarrow b\bar{b}$ decay
width and the $b\rightarrow s\gamma$ branching ratio) are satisfied.   We then compute branching ratios for
$h, H, A$ and $H^+$ and their respective production cross sections and compare with LHC data. We use
SUSHI~\cite{Harlander:2012pb,Harlander:2016hcx} for the
neutral scalars' NNLO production cross sections (computed for the LHC at 13 TeV, of course), and
limit ourselves to the gluon fusion production
process. Other processes could easily be considered, too, but they would not bring anything qualitatively
different from the results presented below. In what follows, we will study first the properties of the
lightest
(SM-like) neutral scalar, and then those of the extra scalar particles predicted in the 2HDM.

\subsection{The SM-like Higgs boson}

With our choice of parameters the lightest CP-even scalar $h$ has a mass of 125 GeV and its properties should
reproduce the LHC results, which indicate a scalar particle behaving very much like what is expected in
the SM. The chosen parameter space -- with $\sin(\beta - \alpha)\geq 0.9$ --  already guarantees that the
couplings of $h$ to the $W$ and $Z$ bosons will be very close to those of SM's.   In addition, given the
form of the quark couplings of $h$, those will also be almost SM-like, as will be those of the couplings
of $h$ to charged leptons, with the exception to the couplings to muons which may be suppressed or
enhanced, depending on the choice of $\tan\beta$, in this muon-specific model.

\begin{figure}
  \centering
  \includegraphics[width=0.7\textwidth]{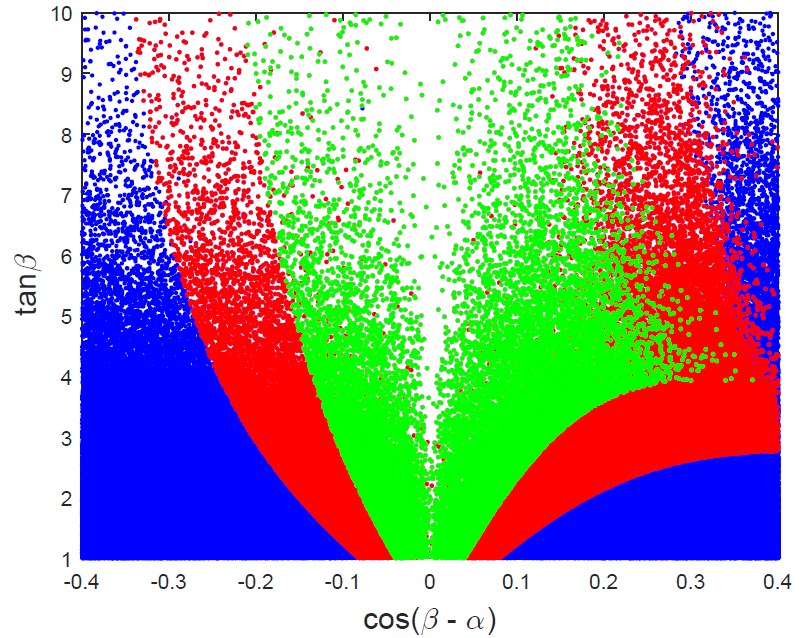}
  \caption{Scatter plot for the variation of $\tan\beta$ as a function of $\cos(\beta-\alpha)$. Blue points
  represent the entirety of the simulated parameter-space (see text). Red (green) points have production
  and decay rates for the 125 GeV neutral scalar within 20\% (10\%) of the expected SM values.
}
  \label{fig:tanb_cosba}
\end{figure}

We first explore the parameter-space allowed in the $\tan\beta-c_{\beta\alpha}$ plane, including LHC data
on Higgs decays.  Since the effects of muons on Higgs branching ratios are negligible, this becomes the
allowed parameter-space of the type I model.  The result is shown in Figure~\ref{fig:tanb_cosba}.
The blue points are all the  points generated within the intervals of variation mentioned above, the
red points require (for gluon production of $h$, and its subsequent decays into $ZZ, WW, \tau\tau, bb,
 \gamma\gamma$) that the cross section times branching ratios be within $20\%$ of their Standard Model
 values, and the green points are within $10\%$ of the Standard Model values. In other words, we compute the
 $\mu_X$ quantites,
\be
\mu_X \,=\, \frac{\sigma^{2HDM} (pp\rightarrow gg\rightarrow h) \,BR^{2HDM}(h \rightarrow X)}
{\sigma^{SM} (pp\rightarrow gg\rightarrow h) \,BR^{SM}(h \rightarrow X)},
\label{eq:muX}
\ee
for the above-mentioned final states, and require that, for all parameters scanned, we have
$|\mu_X - 1|\, <\, 0.2$ (0.1) for the red (green) points. The 10\% requirement on all channels
leads to a parameter space, to a very good degree of approximation, in conformity with current
1-$\sigma$ LHC results. We have not done a full
 $\chi$-squared analysis, since our conclusions won't be affected substantially by doing so.   As expected,
  the resulting parameter-space is close to the usual allowed parameter-space of the type I model. Please
notice that the density of points in this plot has no physical meaning, it is just a consequence of
the fact that certain regions of parameter-space are harder to simulate than others (due to the several
constraints being imposed, both theoretical and experimental).

\begin{figure}
  \centering
  \includegraphics[width=0.7\textwidth]{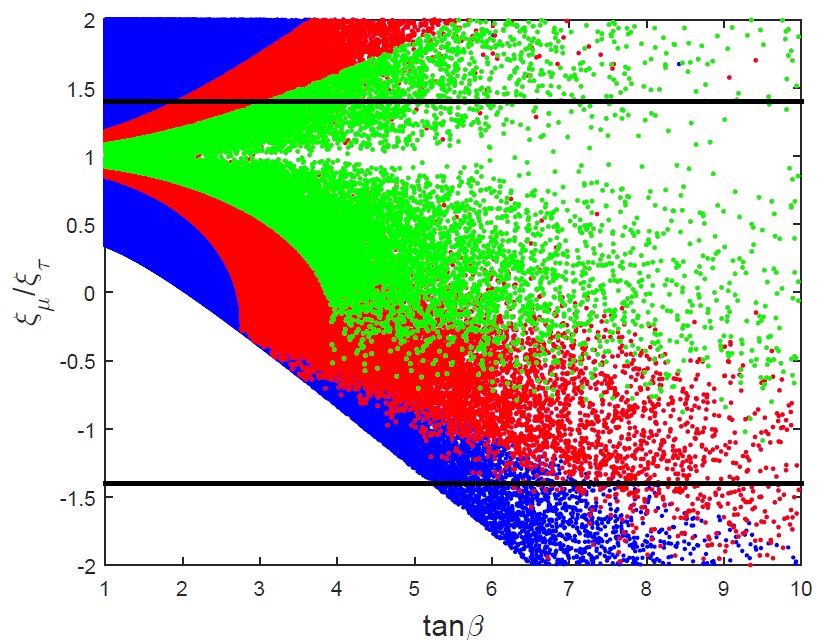}
  \caption{Scatter plot for the variation of $\xi_\mu/\xi_\tau$ as a function of $\tan\beta$. Colour
  conventions as in Figure~\ref{fig:tanb_cosba}.
}
  \label{fig:coupl_ratio_tanb}
\end{figure}

We now turn to predictions obtained for the muon-specific model.
The ratio of the dimuon coupling of the Higgs to the ditau coupling is, as can be seen from Eqs.~\ref{eq:yf}
 and~\ref{eq:ymu},
 only dependent on $\alpha$ and $\beta$, and given $\tan\beta$, the range of $\alpha$ can be
 determined from Figure 1.   Defining $\xi_\mu$ ($\xi_\tau$) as the ratio of the dimuon (ditau)
 coupling of the SM-like Higgs to the Standard Model value, we have, from Eqs.~\ref{eq:yf}
 and~\ref{eq:ymu},
\begin{align}
\xi_\mu &=\,s_{\beta\alpha} - t_\beta c_{\beta\alpha}\,\nonumber
\\
\xi_\tau &=\, s_{\beta\alpha} + \frac{c_{\beta\alpha}}{t_\beta}\,.
\end{align}
If we plot $\xi_\mu/\xi_\tau$ as a function of
 $\tan\beta$ we obtain Figure~\ref{fig:coupl_ratio_tanb}.
 Since the experimental bound on $|\xi_\mu|$ is 1.3 and $\xi_\tau$ must be within approximately $10\%$ of
 unity, the allowed experimental region is between the horizontal lines, corresponding to $|\xi_\mu/\xi_\tau|
 \leq 1.4$ .  One can see that there is a sizeable region in which the dimuon coupling vanishes, as well as a
 region in which it is enhanced.

 As a consequence, the dimuon production rate for the 125 GeV neutral scalar $h$ can be substantially
 suppressed or enhanced. In Figure~3 we show the quantity $\mu_{\mu\mu}$ (defined
 in Eq. \ref{eq:muX} for the final state $X = \mu\mu$)  as a function of $\tan\beta$. Having
 $\mu_{\mu\mu} = 1$ would mean that the dimuon decay of the $h$ would behave exactly like the SM Higgs.
\begin{figure}
  \centering
  \includegraphics[width=0.7\textwidth]{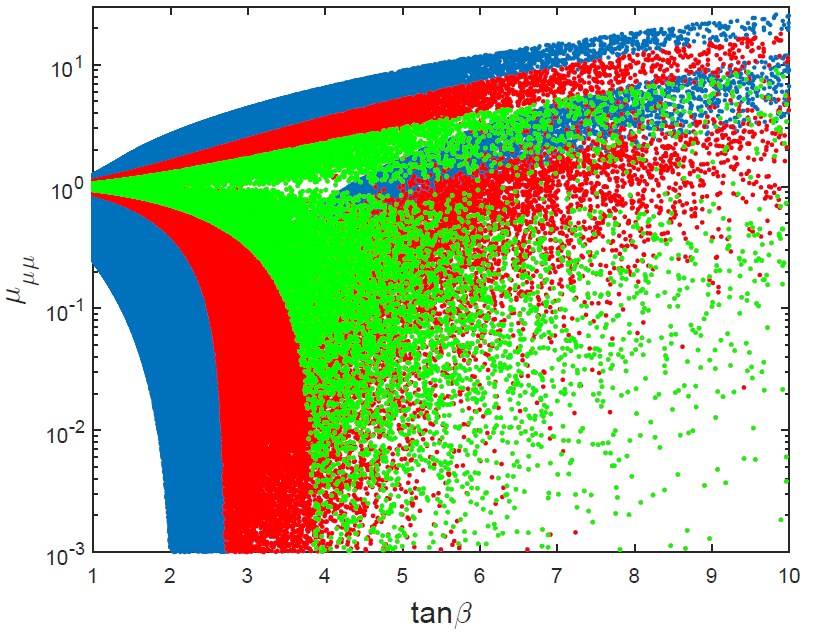}
  \caption{Scatter plot for the variation of the dimuon Higgs rate $\mu_{\mu\mu}$ (as defined in in
  Eq.\ref{eq:muX}) as a function of $\tan\beta$. Colour
  conventions as in Figure~\ref{fig:tanb_cosba}.
}
  \label{fig:mu_rate_tanb}
\end{figure}
 As we see, it is easy to suppress the muon rate to a point where it will never be observed
 at the LHC in the muon-specific model -- but it is also possible to accommodate a muon rate significantly
 larger than the SM expectation.

Since $\xi_Z$, $\xi_\tau$ and $\xi_\mu$ depend only on $\alpha$ and $\beta$, any two will determine the third.    The relation (first noted in Ref. \cite{Dery:2017axi}) is
\begin{equation}
\xi_\mu = \frac{1-\xi_Z \xi_\tau}{\xi_Z - \xi_\tau}.
\end{equation}
This is undefined at $\xi_Z=\xi_\tau=1$, but at that precise point, $\sin(\beta-\alpha)=1$, so $\cos(\beta-\alpha)=0$ and thus $\xi_\mu=1$.   Since we know that $\sin(\beta-\alpha)$ is approximately 1, $\beta - \alpha$ can be written as $\frac{\pi}{2}-\epsilon$.    Expanding in powers of $\epsilon$, one finds
\begin{eqnarray}
\xi_Z &=& 1 - O(\epsilon^2)\cr
\xi_\tau &=& 1 + \epsilon  \cot\beta + O(\epsilon^2)\cr
\xi_\mu &=& 1 - \epsilon \tan\beta  + O(\epsilon^2)
\end{eqnarray}
Note that $\epsilon$ can have either sign.   The fact that $\xi_Z$ is experimentally greater than $0.9$ only
implies that $\epsilon \lessapprox 0.3$.    We see that for moderately large $\tan\beta$, one can easily
suppress the dimuon decay substantially.     In principle, measurements of $\xi_Z$ and $\xi_\tau$ would thus
determine $\xi_\mu$, but given that the uncertainty in these measurements will be at least a few percent for
decades, it will be difficult to be precise.  Note that the model does predict that if the dimuon coupling is
suppressed, then the ditau coupling will be enhanced.

\subsection{Charged Higgs phenomenology}

\begin{figure}[t]
\begin{tabular}{cc}
\includegraphics[height=6cm,angle=0]{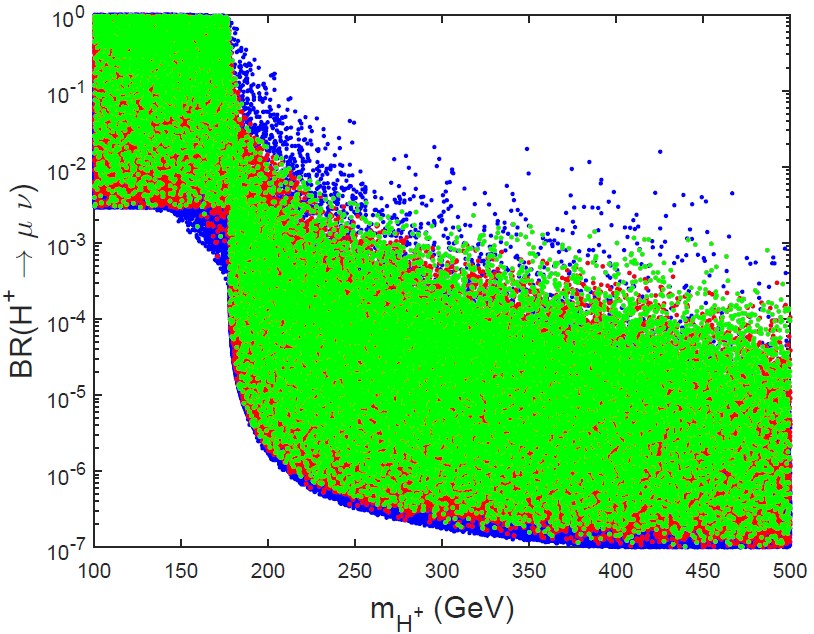}&
\includegraphics[height=6cm,angle=0]{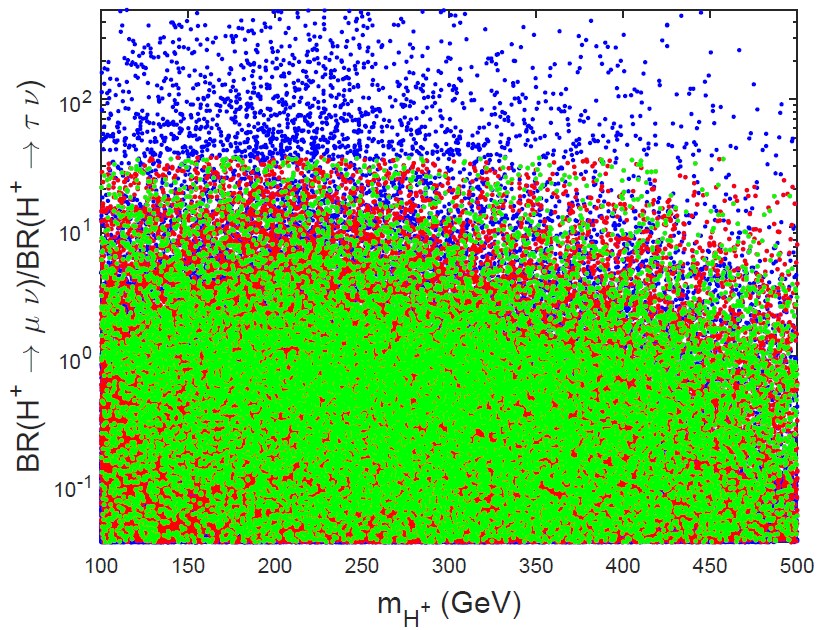}\\
 (a) & (b)
\end{tabular}
  \caption{Scatter plots for (a) the branching ratio for the decay $H^+\rightarrow \mu \nu_\mu$ and (b)
  the ratio of charged Higgs branching ratios to muons and taus, as a function
   of the charged mass $m_{H^+}$. Colour
  conventions as in Figure~\ref{fig:branch_mch}.
}
\label{fig:branch_mch}
\end{figure}

\begin{figure}
  \centering
  \includegraphics[width=0.7\textwidth]{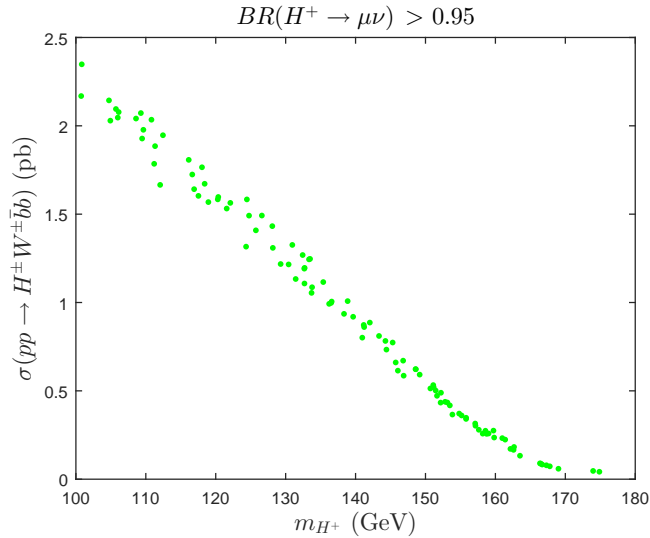}
  \caption{Production cross section for a charged Higgs as a function of its mass for the region of
  parameter space where BR$(H^+\rightarrow \mu \nu_\mu)> 0.95$ and the 125 GeV scalar has production
  and decay rates within 10\% of SM expectations.
}
  \label{fig:sig_mch}
\end{figure}

The charged Higgs can have a completely different phenomenology from that of a Type I model,
since its dominant decay will, for a range of masses and choices of $\tan\beta$, be
$H^\pm \rightarrow \mu\nu_\mu$, instead of the usual
decays into taus. In fact, as can be appreciated from Figure~\ref{fig:branch_mch}, the charged Higgs branching
ratio to muon leptons can be close to unity for a very large range of parameters; and can be substantially
larger than that for tau leptons, even with 125 GeV Higgs rates very close to their SM expectations.
The larger values of BR$(H^+\rightarrow \mu \nu_\mu)$ are clearly obtained for masses of the charged
Higgs inferior to $m_t + m_b \simeq 178$ GeV -- above that mass, the decay channel $H^+\rightarrow tb$
opens up and becomes dominant.
In fact, if one requires the muon decay of the charged Higgs to be dominant, one is left with a
narrow range of masses for which that would be possible. In
Figure~\ref{fig:sig_mch} we show the charged Higgs production cross section as a function of the
scalar mass for a choice of parameters such that the branching ratio of $H^+\rightarrow\mu\nu_\mu$
is $95\%$.   In this case, the muon channel would presumably be the
best discovery channel for the charged Higgs. For this figure we restricted ourselves to points
 for which the 125 GeV neutral scalar has values of the $\mu_X$ ratios within 10\% of SM values,
 and such that $BR(h\rightarrow \mu\mu)/BR^{SM}(h\rightarrow \mu\mu)\leq 1.7$, that is, the current
 LHC bound for muon decays of $h$. The
 main production process for a charged Higgs in this mass range is via production of a top pair, one of
 the top quarks decaying to $Wb$ and the other to $H^+b$. The numbers for the LHC, 13 TeV, cross section were
 obtained from~\cite{Degrande:2016hyf}, duly scaled by appropriate factors of $\tan\beta$. Notice that
 this region of interest -- SM-like muon interactions for $h$, but a subversion of the expectations for
 the charged Higgs phenomenology, with a dominant muon decay channel -- only occurs for values of the
 charged mass below $\sim$ 180 GeV, which justifies the choice of quark-Higgs couplings akin to Type I's. One
 could also choose Type-II-like quark couplings with appropriate choice of quantum numbers, but $b\rightarrow s \gamma$ constraints would automatically
 force the charged mass to be above roughly 580 GeV~\cite{Deschamps:2009rh,Mahmoudi:2009zx,Hermann:2012fc,
Misiak:2015xwa,Misiak:2017bgg,Haller:2018nnx}. Also of note is the fact that the parameter space points
represented in
Figure~\ref{fig:sig_mch} occur for a narrow range of values of $\tan\beta$, namely between $\sim$ 8.97 and
$\sim$ 9.92, which makes them safe from the $B$-physics constraints affecting smaller masses of the
charged scalar for a Type-I 2HDM~\cite{Haller:2018nnx}.

\begin{figure}
  \centering
  \includegraphics[width=0.7\textwidth]{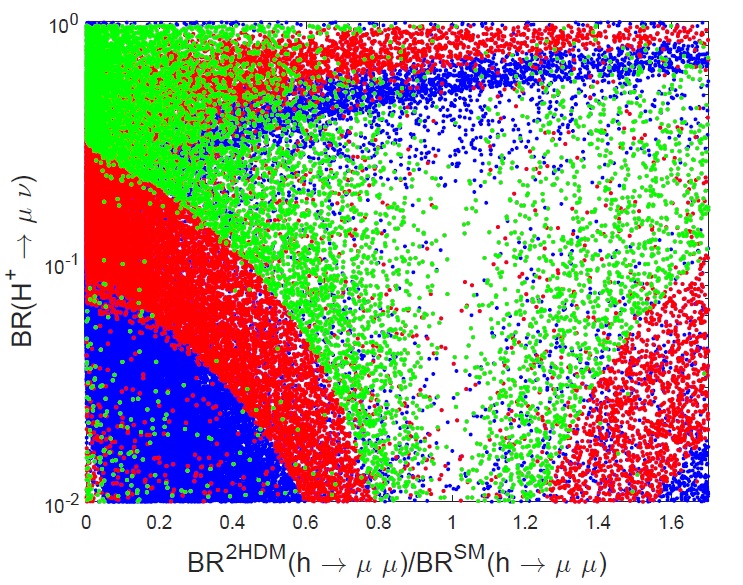}
  \caption{Muonic decay branching ratio of the charged Higgs as a function of the SM-like $h$ branching ratio
  to muons (normalized to the SM expected value). Colour
  conventions as in Figure~\ref{fig:branch_mch}.
}
  \label{fig:brch_mu_brh}
\end{figure}

Finally, one may wonder if the region of parameter space where muonic decays of the charged scalar are
enhanced implies a similar enhancement of the muon branching ratio of $h$. We see, in
Figure~\ref{fig:brch_mu_brh}, that this is not so: the region where the largest values of
 BR$(H^+\rightarrow \mu \nu_\mu)$ occur (even for a scalar $h$ with non-muon production and decay
 rates within 10\% of SM expectations) can in fact occur when the muonic decays of $h$ are highly suppressed
 -- but also, though less likely, when they are enhanced. The explanation, once more, may be found
 in the structure of the muonic Yukawa couplings shown in Eqs.~\ref{eq:yf}--\ref{eq:ych}. We see from
 the latter equation that the muon interactions of the charged scalar are enhanced for high values of
 $\tan\beta$, which simultaneously decreases the tau ones -- thus the branching ratio
  BR$(H^+\rightarrow \mu \nu_\mu)$ will become dominant for high  $\tan\beta$ (and low enough
  mass of the charged, see discussion above for Figure~\ref{fig:branch_mch}). And if $\tan\beta$ is
  of the order of $\sim 1/c_{\beta\alpha}$ then the muonic decays of $h$ will be highly suppressed, as seen
  in Eq.~\ref{eq:ymu}. Larger still $\tan\beta$, however, may actually lead to an enhancement of the magnitude
  of $h$'s muonic Yukawa coupling (though with the opposite sign from the SM expectation). A final observation
  -- in Figure~\ref{fig:brch_mu_brh} we only include values of BR$(h\rightarrow \mu \mu)$ smaller than
  1.7 times the SM Higgs muonic branching ratio, given that that is the current upper bound for that quantity
  stemming from LHC results.

\subsection{Heavier neutral scalars}

\begin{figure}[t]
\begin{tabular}{cc}
\includegraphics[height=6cm,angle=0]{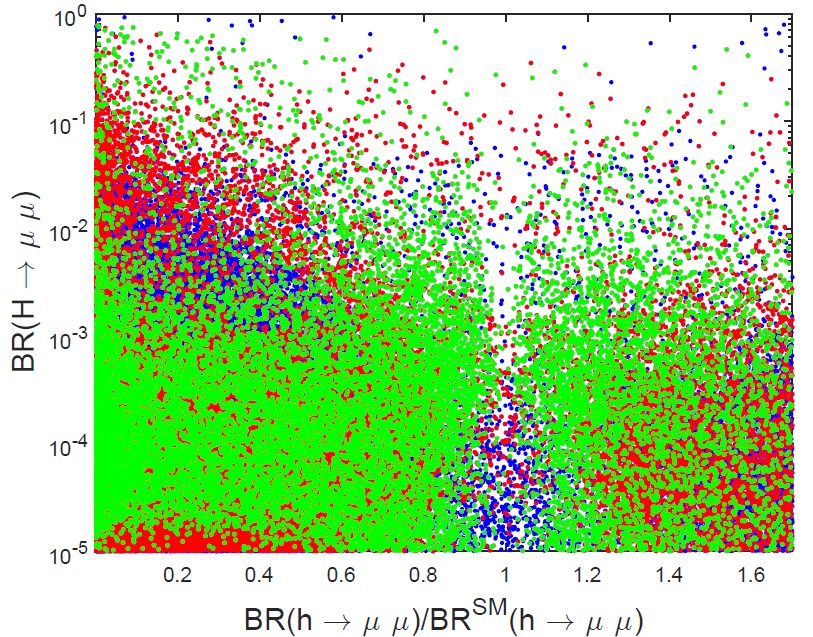}&
\includegraphics[height=6cm,angle=0]{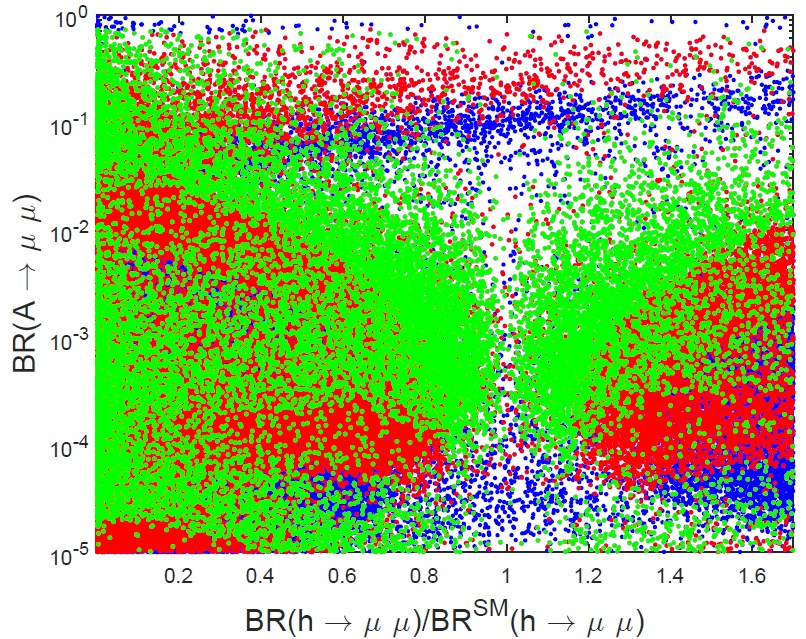}\\
 (a) & (b)
\end{tabular}
  \caption{Scatter plots for the branching ratios for the dimuon decay for (a) the heavier CP-even scalar,
  $H\rightarrow \mu \mu$ and (b) the pseudoscalar,
  $A\rightarrow \mu \mu$, as a function of the SM-like $h$ branching ratio
  to muons (normalized to the SM expected value). Colour
  conventions as in Figure~\ref{fig:branch_mch}.
}
\label{fig:bra}
\end{figure}

\begin{figure}
  \centering
  \includegraphics[width=0.7\textwidth]{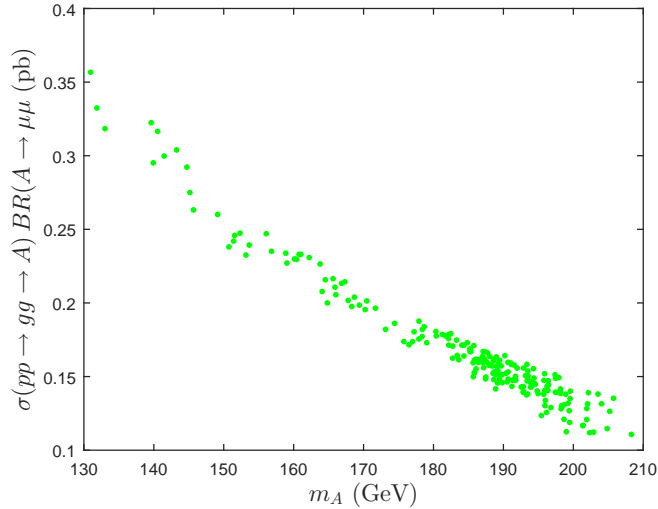}
  \caption{Gluon-gluon production cross section for a pseudoscalar times its branching ratio to muons,
  as a function of its mass for the region of
  parameter space where BR$(A\rightarrow \mu \mu)> 0.5$ and the 125 GeV scalar has production
  and decay rates within 10\% of SM expectations.
}
  \label{fig:sig_A}
\end{figure}

We now consider the heavy neutral scalars in the model. As for $h$ and $H^\pm$, we are interested in
the muon interactions of the heavier CP-even scalar $H$ and the pseudoscalar $A$, so we show, in
Figure~\ref{fig:bra}, the dimuon branching ratios of both $H$ and $A$ as a function of
the muonic branching ratio of the SM-like Higgs (normalized to the expected SM value). We observe that it is
possible to find regions of parameter space for which the muonic decays of both $H$ and $A$ are enhanced and
can even become dominant -- it is easier to accomplish this for the pseudoscalar since its fermionic couplings are independent of $\sin(\beta-\alpha)$ (whereas for $H$ there is a tendency to have Yukawa couplings
suppressed in the regions where $h$ has SM-like interactions, see Eqs.~\ref{eq:yf}--\ref{eq:ymu}). We see then
that there are regions of parameter space for which the decays into muons can become the main decay channel,
raising the possibility of a discovery of these extra scalars by analysing muon pairs produced at the
LHC. In Figure~\ref{fig:sig_A} we show the signal strength for the production of a pseudoscalar $A$ via
gluon-gluon fusion at the LHC at 13 TeV, and its subsequent decay into muon pairs. We chose a region of
parameter space for which the SM-like $h$ has production and decay rates $\mu_X$ within 10\% of the
SM expectation, as explained before; and for which the muonic branching ratio of $A$ is superior to 50\%.
This choice of points includes suppressed $h$ muonic couplings, but also points for which the SM-like
muonic couplings are enhanced. Unlike the charged
Higgs situation, it is not possible to find a region where $h$ has rates within 10\% of SM values {\em and}
BR$(A\rightarrow \mu \mu)>0.95$ -- the maximum value of this branching ratio for this chosen parameter space
is 0.75, though that would have increased had our scan included larger values of $\tan\beta$ (the points shown
in Figure~\ref{fig:sig_A} have values of $\tan\beta$ in the range between $\sim$ 7.7 and 10). And as
occurred for the charged Higgs, only pseudoscalars with lower masses (below $\sim$ 210 GeV) would have a
phenomenology marked by large muonic branching ratios.
Of course, there are many mechanisms for muon pair production at the LHC, and the magnitude of the signal
associated with this pseudoscalar decaying to muons (roughly between 0.1 and 0.4 pb) would likely be drowned in backgrounds.

A similar exercise could be done for the heavier CP-even scalar $H$, but the results are less interesting:
the region of parameter space for which: (a)  $h$ has signal rates within 10\% of its SM values; (b)
its muonic branching ratio is smaller than 1.7 times its SM value; and (c) BR$(H\rightarrow \mu\mu)$ is
dominant, occurs for masses between 130 and 165 GeV. The maximum value of the muonic branching ratio of
$H$ is found to be roughly 0.78, and one obtains $\sigma(pp\rightarrow qq \rightarrow H)\,
BR(H\rightarrow \mu\mu) \lesssim 0.2$ pb, which is even less likely to lead to discovery than the pseudoscalar case.

\subsection{Di-Higgs production}

As noted in the last subsection, backgrounds involving muon pairs can be quite severe.  One can consider di-Higgs production,
through $\bar{q}q \rightarrow Z \rightarrow HA$ or $\bar{q}q \rightarrow W \rightarrow H H^+/A H^+$.
These would lead to spectacular three and four muon events for which the backgrounds would be much smaller.
Note that the dimuon decays of the charged and neutral Higgs are dominant only for a small range of masses
(roughly between 100
and 200 GeV), as shown in the previous subsections, and these three and four muon events require that {\bf both} scalars are in
this small range. Nonetheless, the possibility should be discussed. Assuming that both scalars are produced on-shell, we can use
the results from the 2HDM Benchmark LHC Working Group for the Inert Doublet Model (IDM)-- specifically, their Benchmark Point 5, BP5,
results~\cite{BP5} (see also~\cite{Ilnicka:2015jba,deFlorian:2016spz,Kalinowski:2018ylg}) -- to obtain diHiggs production cross
sections. Given the form of the vertices $Z H A$, $W^\mp H H^\pm$ and $W^\mp A H^\pm$ within the 2HDM, we have~\footnote{Obviously,
if the 2HDM considered was the IDM, we would simply use $\sin(\beta-\alpha) = 1$.} that the production cross sections for $pp\rightarrow Z\rightarrow HA$ and $pp\rightarrow W^\pm\rightarrow H^\pm H$ are $\sin^2(\beta-\alpha)$ times the same processes in the IDM, and the production cross section for $pp\rightarrow W^\pm\rightarrow H^\pm A$ is the same as  $pp\rightarrow W^\pm\rightarrow H^\pm H$ in the IDM.

We wish to ascertain whether in the context of the muon-specific model, where the extra scalars may have very large
muonic decay branching ratios, one can have a very clear 3-- or 4--muon signal. With this in mind, let us provide two
benchmark points (BP) to illustrate the best-case scenario within the model. We require that the 125 GeV neutral scalar has
values of the $\mu_X$ ratios within 10\% of SM values, and that $BR(h\rightarrow \mu\mu)/BR^{SM}(h\rightarrow \mu\mu)\leq 1.7$
(the current LHC bound for muon decays of $h$). With such constraints, we then choose the following BPs:
\begin{itemize}
\item BP1, Maximal neutral muonic branching ratios: we choose parameters to maximize the product
$BR(H\rightarrow \mu\mu) BR(A\rightarrow \mu\mu)$. This will yield the maximal 4--muon signal.
\item BP2, Maximal charged-neutral muonic branching ratios: we choose parameters to maximize the product
$BR(H^+\rightarrow \mu\nu_\mu) BR(H\rightarrow \mu\mu)$. This will yield the maximal 3--muon signal.
\end{itemize}
Investigating our parameter scan detailed in the previous subsections, we find the parameters for each
BP as presented in table~\ref{tab:bps}.
\begin{table}
\begin{center}
\begin{tabular}{rcc} \toprule
Parameters & Benchmark Point 1 & Benchmark Point 2 \\ \midrule
$m_H$ (GeV) &  130.9 &  141.2 \\
$m_A$ (GeV) &  115.2 & 149.6 \\
$m_{H^+}$ (GeV) & 152.0 & 161.3 \\
$\tan\beta$ & 9.2 & 9.4 \\
$\sin(\beta - \alpha)$ & 0.993 & 0.997 \\
$m^2_{12}$ (GeV$^2$) & 1382.9 & 2008.4
\\ \bottomrule
\end{tabular}
\caption{Parameters characterising each of the benchmark points chosen. For both BPs, $m_h = 125$ GeV
and $v = 246$ GeV.
\label{tab:bps}}
\end{center}
\end{table}
We can then read off from the BP5 plots in~\cite{BP5} the rough values for the 13 TeV LHC cross sections
for dihiggs production. For instance, for BP1 we have $\sigma(pp\rightarrow Z\rightarrow HA) \simeq 0.1$ pb and
for BP2 $\sigma(pp\rightarrow W^\pm\rightarrow H^\pm A) \simeq \sigma(pp\rightarrow W^\pm\rightarrow H^\pm A) \simeq
0.04$ pb. Computing the muonic branching ratios of the several scalars and reading off the production cross
sections we obtain the results in table~\ref{tab:res}, with the 3--muon signal $\sigma_{3\mu}$
computed as
\ba
\sigma_{3\mu} &=&\left[\sigma(pp\rightarrow W^\pm\rightarrow H^\pm A) \, BR(A\rightarrow \mu\mu)\frac{}{}\right.  \nonumber \\
 & & +\,\left. \sigma(pp\rightarrow W^\pm\rightarrow H^\pm H) \,BR(H\rightarrow \mu\mu) \right]
 \, BR(H^\pm\rightarrow \mu\nu_\mu)
\ea
and the 4--muon signal $\sigma_{4\mu}$ given by
\be
\sigma_{4\mu} \,=\,\sigma(pp\rightarrow Z\rightarrow H A) \,BR(A\rightarrow \mu\mu)\, BR(H\rightarrow \mu\mu) \,.
\ee
\begin{table}
\begin{center}
\begin{tabular}{rcc} \toprule
  & Benchmark Point 1 & Benchmark Point 2 \\ \midrule
  $BR(H\rightarrow \mu\mu)$ & 0.77 & 0.74 \\
  $BR(A\rightarrow \mu\mu)$ & 0.67 & 0.67 \\
  $BR(H^\pm\rightarrow \mu\nu_\mu)$ & 0.54 & 0.92 \\
  $\sigma(pp\rightarrow Z\rightarrow H A)$ (pb) & $\sim$ 0.1 &  $\sim$ 0.05 \\
  $\sigma(pp\rightarrow W^\pm\rightarrow H^\pm H)$ (pb) & $\sim$ 0.05  & $\sim$ 0.05  \\
  $\sigma(pp\rightarrow W^\pm\rightarrow H^\pm A)$ (pb) & $\sim$ 0.04  & $\sim$ 0.04  \\
 $\sigma_{3\mu}$ (pb) & 0.039 &  0.052 \\
 $\sigma_{4\mu}$ (pb) & 0.051 & 0.0248
\\ \bottomrule
\end{tabular}
\caption{Parameters characterising each of the benchmark points chosen. For both BPs, $m_h = 125$ GeV
and $v = 246$ GeV.
\label{tab:res}}
\end{center}
\end{table}
As before, larger values for $\sigma_{3\mu}$ and $\sigma_{4\mu}$ would be obtained for larger values
of $\tan\beta$ -- for instance, for $\tan\beta = 12.6$, $m_H \sim m_A \sim 130$ GeV and $m_{H^+}\sim 150$ GeV,
one would obtain $\sigma_{3\mu} \simeq 0.069$ and for $\tan\beta = 14.5$, $m_H \sim 133$ GeV, $m_A \sim 116$ GeV
and $m_{H^+}\sim 174$ GeV, one would obtain $\sigma_{4\mu} \simeq 0.085$.

These cross sections are of the order of tens of femtobarns.    There are many analyses of multiple muon events at the LHC in the context of very light scalars, where the scalars are relativistic and thus the muon pairs are collimated (``muon jets").   Here the scalars are not highly relativistic, and thus the opening angle of the muon pairs will be large.  It is possible that a detailed analysis of these processes could exclude some of the model's parameter space, in which both of the heavy scalars are within the $100$ to $200$ GeV range.

\section{Conclusions}

The current experimental value of the dimuon decay of the Higgs boson is $0.5\pm 0.7$ times the
Standard Model expectation.  In this paper, we have questioned the implications of a future result
in which this decay is substantially suppressed.   Since the ditau decay is close to the Standard Model
 expectation, this would mean that the muon and tau must couple to different scalar bosons.   In most
 such models, this results in dangerous tree-level flavor-changing neutral currents, however such
 currents are avoided in the ASY ``muon-specific" 2HDM.    The original ASY analysis hoped to explain
 the g-2 anomaly, but this requires a high degree of fine-tuning.
 Here, we abandon attempting to explain the
 g-2 anomaly, and study the phenomenological implications of the model.

We first study the decays of the 125 GeV Higgs boson.    It is shown that there is a wide range of
parameters in which the dimuon decay rate can be suppressed, even eliminated, without affecting the
other decays.   The ASY model predicts a relationship between the ZZ, ditau and dimuon decays that
 can eventually be tested, although sufficient precision will almost certainly require a Higgs factory.
 The relation shows that a substantial suppression in the dimuon rate does lead to a very small increase
 in the ditau rate.

Since the model is, for all fermions but the muon, close to a type I 2HDM, bounds on charged Higgs masses
are not strong.   For masses below about 175 GeV, we show that the $H^+\rightarrow\mu \nu_\mu$ decay can
dominate the charged Higgs decays, leading to a very different phenomenology.   Above 175 GeV, the $t\bar{b}$ decay becomes accessible and indeed dominant, but for lower masses the muonic decay can dominate over the
tau decay.  For the heavy neutral scalars, one can find substantial regions of parameter space in
which the dimuon decay dominates, especially for the pseudoscalar Higgs.    This also occurs for fairly
light masses.    These decays, given the large number of muon pairs at the LHC, may be swamped by
other processes.   On the other hand, if two of the heavy scalars are in a similar mass range, then multimuon events could provide a distinctive signature.

One could imagine using the same mechanism of the ASY model in the quark sector, to enhance or
suppress couplings of $h$ to the second or first generation of quarks, the measurements of which
have not yet been achieved. One could imagine that probing the Higgs-charm coupling might be
possible indirectly, via interference effects in the gluon-gluon fusion cross section, or in
the diphoton decay width. A strong enhancement of the coupling of $h$ to charm quarks (a ``{\em charming
Higgs}") could then be ruled out. Likewise one could suppress this coupling ( ``{\em charmless
Higgs}"), though the exclusion of that possibility seems unlikely within the expected lifetime of the
LHC. Similar ideas could be explored {\em vis-a-vis} couplings of $h$ to $s$, $d$ or $u$ quarks. One might
even consider enhancing/suppressing the third generation quark couplings, if increased precision in their
measurements showed deviations  unable to be reproduced by regular 2HDMs. However, the ASY mechanism runs
into a serious obstacle if one tries to extend it to the quark sector -- since it would require that for
a given generation one of the quark left doublets and the corresponding quark right singlet transform
differently from the other two generations, attempting to reproducing the ASY mechanism in the
quark sector would result in an unphysical CKM matrix -- namely, one would find that the CKM matrix would be
block diagonal, thus contradicting experimental confirmation of all its elements being non-zero. This
conundrum might be solved with a more complicated scalar and/or quark sector, but that goes beyond the scope
of the current work.

\subsubsection*{Acknowledgments}
PF is supported in part by a CERN grant CERN/FIS-PAR/0002/2017, an FCT
grant PTDC/FIS-PAR/31000/2017, by the CFTC-UL strategic project
UID/FIS/00618/2019 and by the NSC, Poland, HARMONIA UMO-2015/18/M/ST2/00518.
The work of MS was supported by the National Science Foundation under Grant PHY-1819575.  We thank Igor Ivanov for clarifying the symmetry group of the model.

\end{document}